# SOLVING THE GENERAL CASE OF RANK-3 MAKER-BREAKER GAMES IN POLYNOMIAL TIME

LEAR BAHACK

lear.bahack@gmail.com

Abstract. A rank-3 Maker-Breaker game is played on a hypergraph in which all hyperedges are sets of at most 3 vertices. The two players of the game, called Maker and Breaker, move alternately. On his turn, maker chooses a vertex to be withdrawn from all hyperedges, while Breaker on her turn chooses a vertex and delete all the hyperedges containing that vertex. Maker wins when by the end of his turn some hyperedge is completely covered, i.e. the last remaining vertex of that hyperedge is withdrawn. Breaker wins when by the end of her turn, all hyperedges have been deleted.

Solving a Maker-Breaker game is the computational problem of choosing an optimal move, or equivalently, deciding which player has a winning strategy in a configuration. The complexity of solving two degenerate cases of rank-3 games has been proven before to be polynomial. In this paper, we show that the general case of rank-3 Maker-Breaker games is also solvable in polynomial time.

Algorithms in Graphs; Combinatorial Games; Computational Complexity; Complexity of Games; Game Theory; Hypergraphs; Maker-Breaker; POS-CNF Problem.

## 1. INTRODUCTION

The Maker-Breaker game is a positional game, firstly introduced in the classical paper[4] that proves sufficient (but unnecessary) winning conditions for both Maker and Breaker. While Maker-Breaker games are mostly researched in extremal combinatorics and similar contexts, a few important papers[1, 2, 9, 10] deal with the computational complexity of solving the games.

Shaefer[10] was the first to show a (polynomial) reduction from solving rank-11 Maker-Breaker games to solving general TQBF problems, thus proving that solving the general Maker-Breaker game is PSPACE-complete. This result is of the highest significance in the field of games complexity, since for many combinatorial games establishing a reduction to the Maker-Breaker game is far more natural than to classical PSPACE-hard computational problems. The Maker-Breaker game, thus, is a major tool in the research of other games complexity. We note that Shaefer and others often use an equivalent formulation of the Maker-Breaker game, in terms of positive CNF formulas, instead of the hypergraphs formulation used here.

Our two-parts work closes the existing gap between rank-6 games (that are PSPACE-complete to solve[9]) and two degenerate cases of rank-3 games that are solved in polynomial time[8, 7]. In this paper, we solve the (general case) rank-3 game in polynomial time, and in the complementing paper[1] we prove that solving rank-4 games is a PSPACE-complete problem.

1.1. **Related Works.** Lutfar and Watson[8] solved (in polynomial time) rank-3 games in which every edge of the hypergraph contains a unique vertex that is not shared by other edges. Their technique is strongly dependent on those unique vertices, and cannot be generalized. However, their algorithms is not merely in P, but in the complexity class L.

Kutz[7] solved (in polynomial time) rank-3 games in which distinct edges may share at most one vertex in common. While this case is also a degenerate case, and dealing with pairs of edges sharing two vertices imposes great challenges, we have taken a similar course of thought. In fact, we believe Kutz have been close to solving the general case. However, his style of branching into numerous cases, makes any adaptation to the general case too difficult (at least for humans).

1.2. **Our Technique and Structure of the Paper.** In our work, we develop some tools and general theory of rank-3 games. We manage to keep the number of separate cases in our proof relatively small, by systematically applying our theoretical tools to analyze the structure of the games considered. Yet the reader is warned that this paper is rather technical.

We define a shortened version of the Maker-Breaker game, by introducing additional winning conditions for Maker. The new conditions are sufficient to ensure (in optimal play) the winning of Maker in the original version, and thus the shortened version is equivalent. We call it a shortened version, because it essentially enables Maker-Breaker games to end sooner, making the game shorter.

The polynomial solving algorithm provided is a standard mini-max algorithm, based on our main theoretical theorem that bounds the maximal number of moves (in optimal play) in the shortened version. The theorem's proof is by contradiction: we assume there is a game in which Maker has a winning strategy but winning takes too many moves, and systematically conclude conflicting properties of the game's hypergraph.

In section 2 we introduce terminology, sufficient winning conditions (of Maker), our shortened version of the game, and numerous lemmata. In section 3 we heavily analyze shortened games, and bound their length (number of moves it takes for Maker to win). In section 4 we provide the polynomial time algorithm for solving general rank-3 games, and analyze its running time. Section 5 closes the paper with three further research ideas.

## 2. Definitions and Lemmata

2.1. **Hypergraphs.** A rank-3 *hypergraph* $H = (V, E)$ consists of a set of *vertices* $V$ and a set of *edges* (hyperedges) $E \subseteq \mathcal{P}(V)$ containing subsets of $V$ of cardinality at most 3. An edge of cardinality $k$ is called a $k$-edge, or singleton for $k = 1$, and the empty edge for $k = 0$. In this paper all hypergraphs are assumed to be finite, and edges are usually (see lemma 9) assumed not to be subsets of other edges.

Two 3-edges $e_1, e_2 \in E$ are said to be *overlapping* if $e = e_1 \cap e_2$ is of cardinality two, and (assuming $e \notin E$) we regard $e$ as a *virtual 2-edge* of the hypergraph, generated by $e_1, e_2$. A 3-edge that is not overlapping any other edge is called a *regular* edge. A vertex included in 3-edges only, and not included in any virtual 2-edge, is called a *regular vertex*.

For disjoint sets $X = \{x_1, \ldots x_n\}, Y = \{y_1, \ldots y_m\} \subseteq V$ we define:

$$E_{y_1, \ldots y_m}^{x_1, \ldots x_n} = \{e \backslash X | e \in E \text{ and } e \cap Y = \emptyset\}$$

and:

$$H_{y_1, \ldots y_m}^{x_1, \ldots x_n} = \left( V \backslash (X \cup Y), E_{y_1, \ldots y_m}^{x_1, \ldots x_n} \right)$$

so $H_{y_1, \ldots y_m}^{x_1, \ldots x_n}$ is essentially the hypergraph we get from $H$ after deleting all edges containing any vertex of $Y$, withdrawing all vertices of $X$ from the existing edges (without deleting them, not even the empty edge $\emptyset$), and removing the vertices of $X \cup Y$ from the vertices list.

2.2. **Types of Paths.** A *path* is a sequence of edges $(e_1, \ldots, e_n)$ s.t. consecutive edges share at least one vertex in common: $\forall 1 \leq i < n : e_i \cap e_{i+1} \neq \emptyset$. The edges $e_1, e_n$ are called the *ends* of the path, and the remaining edges are the *intermediate* edges.

A *cycle* is a path $(e_1, \ldots, e_n)$, $n \geq 3$, where $e_1 \cap e_n \neq \emptyset$ and $e_1, e_n$ are considered intermediate consecutive edges. We say a vertex $u$ has a cycle, when there exists a cycle which is a $u - u$ path, i.e. $u \in e_1 \cap e_n$. Note that $u$ might have no cycles, despite belonging to an edge of a certain cycle (but to neither of its neighboring edges).

A path (or a cycle) is called *linear* when consecutive edges share exactly one vertex in common, and non consecutive edges are disjoint. A path (or a cycle) is called *3-uniform* if it consists of 3-edges only. A linear path (or a cycle) whose intermediate edges are all regular is called a *regular path* (or a cycle).

Two vertices $u, v$ are said to be *connected* if there exists a $u - v$ path (i.e. a path $(e_1, \ldots, e_n)$ s.t. $u \in e_1, v \in e_n$), *linearly connected* if there exists a $u - v$ linear path, and *regularly connected* if there exists a $u - v$ regular path.[1]

A $u - v$ linear path $(e_1, \ldots, e_n)$ admits a unique natural order of its participating vertices, starting with $u$ and ending with $v$, such that the vertex in the intersection $e_i \cap e_{i+1}$ precedes the other vertices of $e_{i+1}$ and exceeds the other vertices of $e_i$.

For a $u - v$ linear 3-uniform path $\alpha = (e_1, \ldots, e_n)$ s.t. $u$ is included only on the first edge $e_1$ and $v$ is only on the last edge $e_n$, we let $\alpha^{u,v}$ denote the linear path $(e_1 \backslash \{u\}, \ldots, e_n \backslash \{v\})$. If vertex $u$ has a linear 3-uniform cycle $\gamma = (e_1, \ldots, e_n)$, we let $\gamma^u$ denote the linear path $(e_1 \backslash \{u\}, \ldots, e_n \backslash \{u\})$.

---

[1]While connectivity is an equivalence relation, linear and regular connectivity are not (transitivity fails). Coping with this non-transitivity is the main challenge in our research.

2.3. **Cutters and Partial Orders.** Linear and regular paths are crucial for analyzing rank-3 games, and understanding their nuances is important.

If $u, v, c$ are distinct vertices s.t. $u, v$ are not regularly connected in $H_c$, we say that $c$ is a $u - v$ *cutter* in the hypergraph $H$.

For a fixed source vertex $s$, we define the relations $\preceq_s$ and $\prec_s$ as follows: $u \preceq_s v$ iff every linear $s - v$ path contains $u$, and $u \prec_s v$ iff $u \preceq_s v$ but not $v \preceq_s u$.

**Fact 1.** *The vertices set $V$ equipped with the relation $\prec_s$ of a rank-3 hypergraph $H = (V, H)$ is a (strict) partially ordered set.*

*Proof.* Directly by the definition. □

*Remark* 2. A path $(e)$ consisting of a single edge $e$ is regular by our definition, even if $e$ is not a regular edge.

*Remark* 3. The only cases where a path $(e, t_1, \ldots, t_n, f)$ with $n \geq 0$ intermediate regular edges is not a regular path (essentially because it is not linear) are when $n = 0$ and $e, f$ overlap, or $n \geq 1$ and $e$ is not disjoint from an edge other than $t_1$, or $n \geq 0$ and $f$ is not disjoint from an edge other than $t_n$.

*Remark* 4. If $u, v$ are not regularly connected in $H$, then by definition all vertices of $H$ are $u - v$ cutters.

*Remark* 5. If $(e_1, t_1, \ldots, t_n, f)$ and $(e_2, t_1, \ldots, t_n, f)$ are regular paths and $e = e_1 \cap e_2 \notin E$ is a virtual 2-edge of the overlapping 3-edges $e_1, e_2$ then we think of $(e, t_1, \ldots, t_n, f)$ as a (one-sided) virtual regular path. Two-sided virtual regular paths are defined similarly.

*Remark* 6. In case $e_1, e_2$ overlap, $e = e_1 \cap e_2 \notin E$, and $(e, t_1, \ldots, t_n, f)$ is a non-virtual regular path in the hypergraph $(V, E \cup \{e\})$, then $(e, t_1, \ldots, t_n, f)$ is a virtual path in $(V, E)$ if the two vertices of $(e_1 \cup e_2) \setminus e$ do not belong to any of the edges of $(e, t_1, \ldots, t_n, f)$.

*Remark* 7. It is possible for $u \neq v$ that both $u \preceq_s v, v \preceq_s u$ are true.

**Lemma 8.** *Let $\alpha = (e_1, \ldots, e_n)$ be a $u - v$ regular path and $\beta = (f_1, \ldots f_m)$ be a $w - v$ regular path, $u \in e_1, v \in e_n \cap f_m, w \in f_1$. If $e_n \cap f_m = \{v\}$ and $e_1, f_1$ do not overlap, there exists a $u - w$ linear path. In case $e_n, f_m$ are also regular, there exists a $u - w$ regular path.*

*Proof.* We may assume $e_1 \cap f_1 = \emptyset$, or else $e_1, f_1$ form a regular path (possibly of length 1, if $e_1 = f$). Let $e_k$ be the first edge of $\alpha$ containing a vertex that is on (any of the edges of) the path $\beta$, and let $f_l$ be the first edge of $\beta$ sharing a common vertex with $e_k$. Then a possible path meeting the requirements is $(e_1, \ldots, e_k, f_l, \ldots f_1)$, which is the concatenation of the first portion of $\alpha$ with the (reversed order of) the first portion of $\beta$. □

2.4. **The Maker-Breaker Game.** A rank-3 Maker-Breaker game[2] $G = (V, E, P)$ is any rank-3 hypergraph $(V, E)$ together with a designated player whose turn is to move next $P \in \{M, B\}$, where $M$ and $B$ stands for Maker and Breaker. We view an intermediate configuration that occurs while playing a game, as a game by itself. For $G = (V \setminus \{x\}, E, P)$ and $x \in V$ we define $G^x = (V, E^x, B)$, and for $G = (V, E, B)$ and $y \in V$ we define $G_y = (V \setminus \{y\}, E_y, M)$

The rules, informally described in the abstract, are defined by setting the *winning and losing endgames* (terminal games, with no further possible moves), and the set of legal moves in each non-terminal game. We sympathize with Maker, so a winning endgame is a win for Maker, and a losing endgame is a win for Breaker.

A game $G = (V, E, B)$ is a winning endgame if $E$ contains the empty edge $\emptyset$ , and $G = (V, E, M)$ is a losing endgame if $E = \emptyset$. For a non-terminal game $G = (V, E, M)$ the legal moves are all games of the form $G^x$ for $x \in V$, and for a non terminal $G = (V, E, B)$ the legal moves are all games $G_y$ for $y \in V$. Essentially, a winning endgame is when Maker has reached his winning condition (taking all the vertices of an original edge), and a losing game is when Maker has continuously failed to do, until he run out of moves and loses by default. Maker is called this way because he is trying to make the winning condition, and his opponent is called Breaker because she is trying to break all of his possible winning conditions.

For analysis purposes, we are interested in the game resulting from switching the identity of the player moving next. For $G = (V, E, M)$ and $F = (V, E, B)$ we let $G^* = F$ and $F_* = G$. Similarly,

$$G^x_* = (G^x)_* = (V \setminus \{x\}, E^x, M) \quad , \quad F^*_y = (F_y)^* = (V \setminus \{y\}, E_y, B)$$

In a sense, $*$ serves as a passing move, i.e. a move in which the player takes no variable.[3]

2.5. **Optimal Moves and Winning Depth of Games.** As a perfect information 2-players alternating turns game, in each rank-3 Maker-Breaker game (exactly) one of the payers has a winning strategy. We extend the definition of winning and losing games beyond terminal games, by calling a game $G$ a winning (resp. losing) game if Maker (resp. Breaker) can force a win.

We define the *depth $dep(G)$* of a winning game $G$ recursively, by setting $dep(G) = 0$ for winning endgames. For a non terminal $G = (V, E, M)$:

$$dep(G) = 1 + \min_{x \in V} dep(G^x)$$

and for a non-terminal $G = (V, E, B)$:

$$dep(G) = 1 + \max_{y \in V} dep(G_y)$$

---

[2]Or in alternative form $G = (H, P)$, for a rank-3 hypergraph $H$.

[3]Clearly, in Maker-Breaker game a player can only benefit from making a move, so it is even possible to introduce passing as additional legal move, and get an equivalent game.

essentially the depth is the smallest (resp. largest) number of move Maker (resp. breaker) can guarantee until a winning endgame is eventually reached. The depth of a losing game is naturally defined to be $\infty$.

A move $x$ of Maker is considered *optimal* for $G = (V, E, M)$ when

$$dep\,(G^x) = \min_{u \in V} dep\,(G^u)$$

and similarly, a move $y$ of Breaker is optimal for $G = (V, E, B)$ when

$$dep\,(G_y) = \max_{u \in V} dep\,(G_u)$$

Note that our notion of an optimal move is more narrow than usual: a move $x_1$ of Maker can be part of a winning strategy (namely, $G^{x_1}$ be a winning game) and yet not optimal, in case there exists a move $x_2$ s.t. $dep\,(G^{x_1}) > dep\,(G^{x_2})$.

**Lemma 9.** *Assume $G = (V, E, P)$, $G' = (V, E \setminus \{f\}, P)$ where $e, f \in E$ are such that $e \subsetneq f$. Any sequence of moves is winning in $G$ iff it is winning in $G'$.*

*Proof.* After any move played, by any player, it is always the case that any edge of (the resulting sub-game of) $G'$ is also an edge of (the sub-game of) $G$. If a sequence of moves is winning in $G'$, the resulting endgame (of $G'$) contains $\emptyset$ as an edge, thus the resulting sub-game of $G$ also contains $\emptyset$, and is therefore a winning endgame.

For the opposite direction, let $e', f'$ be the resulting edges $e, f$ after all vertices played by Maker are removed. The only potential case of a sequence of moves that is winning in $G$ but not in $G'$ is where $f' = \emptyset$, and non of the vertices of $f$ have been played by Breaker (otherwise the edge $f'$ would have been deleted). $e'$ has not been deleted either (any Breaker move deleting $e$ would be also deleting $f$), and obviously $e' \subseteq f'$. Therefore it must be that $e' = \emptyset$ and the sequence is winning in $G'$ too.                              $\square$

Based on the above lemma, we can safely assume that distinct edges of our Maker-Breaker games are never subsets of each other. Note that redundant edges (edges for which a proper subset is also an edge) can be detected and removed in polynomial time easily. Following a Maker's move $x$ which causes an edge $e$ (containing $x$) to become a proper subset $e \setminus \{x\}$ of another edge $f$, we shall remove $f$ from the hypergraph.

2.6. **Hot Paths.** We say that three (distinct) vertices $u, v, w$ in a game $G = (V, E, P)$ form a *v-shape* if the two edges $\{u, v\}, \{v, w\}$ are in $E$.

**Lemma 10.** *If $G = (V, E, M)$ contains a v-shape, it must be a winning game.*

*Proof.* Assume $\{u, v\}, \{v, w\} \in E$. After maker takes $v$, there are two singletons $\{u\}, \{w\}$ in $G^v$. Assume wlog that Breaker responds with $y \neq u$, and then $G_y^v$ still contains $\{u\}$, so $G_y^{v,u}$ is a winning endgame.

Interestingly, the converse (with the right formulation) is also true.        $\square$

**Lemma 11.** *Any optimal winning sequence of moves, starting from a winning game in which there are no edges of cardinality less than 2, must go through a game $G = (V, E, M)$ containing a v-shape.*

*Proof.* The length of any such a sequence is of at least 4 games (or 3 moves), because Maker has to make at least two moves in order to withdraw all vertices of an edge of size at least 2. Let the last 4 games of the sequence be $G, G^v, G^v_u, G^{v,w}_u$.

In the last move, Maker withdraws $w$ from $\{w\}$, and since he plays optimally, he couldn't have done it earlier, so $\{w\}$ is an edge of $G^v_u$ but not of $G$, which means $\{v, w\}$ is an edge of $G$.

Since Breaker is playing optimally too, taking $w$ instead of $u$ wouldn't let her make the game last longer, so inevitably $\{u, v\}$ is also an edge of $G$. $\qquad\square$

We can generalize the v-shape to get wider winning conditions for games in which Maker moves next.

**Lemma 12.** *If $G = (V, E, M)$ contains a linear path:*

$$(\{y_1, x_1\}, \{x_1, y_2, x_2\}, \{x_2, y_3, x_3\}, \dots, \{x_{n-2}, y_{n-1}, x_{n-1}\}, \{x_{n-1}, x_n\})$$

*that connects a pair of 2-edges $\{y_1, x_1\}, \{x_{n-1}, x_n\}$ by a subsequence[4] of intermediate 3-edges, then $G$ is a winning game.*

*Proof.* Consider the following sequence of moves, starting and ending with Maker: $x_1, y_1, x_2, y_2, \dots, y_{n-1}, x_n$. All of Breaker's moves $y_i$, $1 \le i < n$, are essential to remove the resulting singleton edge $\{y_i\}$, so any alternative move for Breaker enables Maker to win by the following turn. Thus Maker has a winning strategy in $G$. $\qquad\square$

And yet, we can push it even further.

**Lemma 13.** *Lemma 12 remains valid when we allow one (or both) of the edges $\{y_1, x_1\}$, $\{x_{n-1}, x_n\}$ at the ends of the path to be virtual, as long as the two overlapping 3-edges creating the virtual edge are disjoint from the remaining edges of the path.*

*Proof.* Assume $\{y_1, x_1\}$ is virtual, so there are two distinct vertices $w_1, w_2$, different from all of the $x_i$-s and $y_i$-s, such that instead of the virtual $\{y_1, x_1\}$, there are two overlapping edges $\{w_1, y_1, x_1\}, \{w_2, y_1, x_1\} \in E$. The case where $\{x_{n-1}, x_n\}$ is virtual, or both ends are virtual, is proven similarly.

The sequence of moves $x_1, y_1, x_2, y_2, \dots, y_{n-1}, x_n$ still applies: following Maker's first move at $x_1$, the resulting game $G^{x_1}$ contains a v-shape $\{w_1, y_1\}, \{w_2, y_1\}$ that must be destroyed by the following move of Breaker if she can force a win. Breaker might respond with $y_1$, as in the sequence of moves of lemma 4, or alternatively with either of $w_1, w_2$. Then Maker plays at $x_2$, and the proof continues the same as of lemma 12. $\qquad\square$

A linear path connecting two distinct virtual/proper 2-edges through intermediate 3-edges only (as in lemmata 12,13) is said to be a *hot path*. In case the path is regular, it is said to be a *regular hot path*.

Since hot paths are by definition linear paths, and linear connectivity is not transitive, the trivial generalization of classical graph traversal algorithms[5] to hypergraphs cannot help us decide, in polynomial time, whether a pair of virtual/proper 2-edges is connected by a hot path. The recent interesting paper [5] provide such a (polynomial) algorithm, and yet is unneeded for our purpose: we only wish to decide whether a hypergraph contains any hot path, between any possible 2-edges.

---

[4]An empty subsequence of 3-edges results in the v-shape case (lemma 10).

[5]See the BFS and DFS algorithms in [3], chapter 22.

**Lemma 14.** *A hypergraph $H = (V, E)$ contains a hot path iff it contains a regular hot path.*

*Proof.* Consider a hot path $(e, t_1, \ldots, t_n, f)$ of the shortest possible length in $H$, and assume for contradiction that it is not regular. Let $t_i = \{u, v, w\}$ be a non regular 3-edge of the hot path, and let $\{u\}$ and $\{w\}$ be the intersections of $t_i$ with its preceding and succeeding edges, respectively. Let $t_i'$ be an edge overlapping $t_i$, and $t = t_i \cap t_i'$ be the virtual 2-edge they form.

At lest one of $u, w$ is in $t$. If $u \in t$ then $(e, t_1, \ldots, t_{i-1}, t)$ is a hot path, and if $w \in t$ then $(t, t_{i+1}, \ldots, t_n, f)$ is. In both cases we got a (strictly) shorter hot path, contradicting our assumption. $\square$

**Corollary 15.** *The complexity of deciding whether a hypergraph contains a hot path is polynomial.*

*Proof.* Based on lemma 14, the algorithm is almost trivial: compute the sub-hypergraph of all regular 3-edges, and the set of virtual and proper 2-edges, then use a linear time (hypergraph trivial generalization of a) graph traversal algorithm[3] to compute the connected components of the sub-hypergraph, and check whether a pair of 2-edges is touching the same connected component.

The weak transitivity of lemma 8 is enough to ensure the sub-hypergraph indeed decomposes to connected components. However, in case of virtual 2-edges, a pair touching the same connected component is only a candidate for a hot path: say $e$ is a virtual 2-edge of the overlapping 3-edges $e_1 = e \cup \{u_1\}$, $e_2 = e \cup \{u_2\}$, then we need to make sure that after removing $u_1, u_2$ from the connected component, connectivity of the pair of 2-edges remains. $\square$

Another immediate corollary of lemmata 12,13 is:

**Corollary 16.** *If all of Breaker's moves in a game $G = (V, E, B)$ result in games $G_y$ containing regular hot paths, $G$ is a winning game.*

2.7. **Shortened Rank-3 Maker-Breaker Games.** When for each game in a family of non terminal games, we can verify in polynomial time that it is a winning game, it makes sense to consider all games of this family as winning endgames, despite the fact that they are not really terminal games. Consider the following two families of games, that are promised to be winning games by lemma 5 and corollary 8.

FAM1:     Games where Maker is moving first, containing a singleton edge or a hot path.

FAM2:     Games where Breaker is moving first, such that all of her possible moves result with games of FAM1.

For the sake of completeness of notation, we regard original terminal winning games as

FAM0:    Games where Breaker is moving first, containing the empty set $\emptyset$ as an edge.

We define a shortened (yet equivalent) version of the rank-3 Maker-Breaker game, by letting the set of shortened winning endgames include FAM1 and FAM2 in addition to FAM0.

**Fact 17.** *The complexity of deciding the set of winning shortened endgames is still polynomial.*

*Proof.* For deciding FAM1, check whether there is an edge of cardinality 1, and if not, use the algorithm of corollary 7 to check if there is a hot path.

For FAM2, apply the algorithm of the first family over all possible first moves of Breaker. Repeating a polynomial algorithm for polynomially many times, still yields a polynomial algorithm. □

The shortened-depth $sdep\,(G)$ is defined similarly to $dep\,(G)$, by setting $sdep\,(G) = 0$ for shortened winning endgames, using the same recursive rules for non terminal winning games, and setting $sdep\,(G) = \infty$ for losing games. From this point forward, optimality of moves (see §2.5) is w.r.t the shortened depth.

**Lemma 18.** *Any winning sequence of (shortened) optimal moves starting with a non terminal shortened game, must end with a shortened endgame in FAM2.*

*Proof.* The sequence cannot end with a FAM0 game, since first it would have to go through a game containing a singleton edge, which is part of FAM1 and is therefore terminal.

However, reaching a FAM1 endgame is also impossible: the preceding game of a FAM1 game must be a (terminal) FAM2 game. This is due to the optimality of the last move. □

**Corollary 19.** *If Breaker moves first in $G$, $sdep\,(G)$ is even, and if Maker moves first then either $G$ is an endgame of the first family (i.e. $sdep\,(G) = 0$) or $sdep\,(G)$ is odd.*

*Proof.* Immediately follows from lemma 10, by induction over $sdep\,(G)$. □

**Lemma 20.** *If $G_1 = (V_1, E_1, M) \in FAM1$ and $G_2 = (V_2, E_2, M)$ is a game containing all singletons and hot paths of $G_1$, then $G_2 \in FAM1$.*

*If $G_1 = (V_1, E_1, B) \in FAM2$ and $G_2 = (V_2, E_2, B) \notin FAM0$ is a game containing all singletons and hot paths of $G_1$, then $G_2 \in FAM2$.*

*Proof.* In the first case, $G_1 \in FAM1$ iff there exists a singleton or a hot path in $G_1$. Thus there exists a singleton or a hot path in $G_2$, meaning $G_2 \in FAM1$.

In the second case, let $y_2 \in V_2$ be an arbitrary possible move of Breaker, and we prove that $(G_2)_{y_2} \in FAM1$. Let $y_1 \in V_1$ be the same vertex $y_2$ as long as $y_2 \in V_1$, and an arbitrary vertex otherwise. By definition of FAM2 we know that $(G_1)_{y_1} \in FAM1$, and since every singleton and hot path of $(G_1)_{y_1}$ must be present in $(G_2)_{y_2}$, the first case of the lemma is applied. □

**Lemma 21.** *If $G_*^u$ is not a winning shortened endgame, then $u$ is a regular vertex in $G$ and for any vertex $v \neq u$ there exists a $u - v$ cutter $c$ in $G$.*

*Reminder.* Maker moves first in $G_*^u$ (see §2.4) and hence $G_*^u \notin$ FAM2 by definition.

*Proof.* $u$ must be a regular vertex in $G$, or else $G_*^u$ contains a singleton (if $u$ belongs to a 2-edge of $G$) or a v-shape (if $u$ belongs to a virtual 2-edge of $G$). Assume for contradiction that there are no $u - v$ cutters in $G$, and $G_*^u \notin$ FAM1.

For every $u - v$ regular path $\alpha = (e_1, \ldots, e_n)$, $u \in e_1, v \in e_n$, since the two vertices of $e_1 \setminus \{u\}$ are not cutters, there must exist a $u - v$ regular path $\beta = (f_1, \ldots, f_m)$, $u \in f_1$, $v \in f_m$ s.t. the edges $e_1, f_1$ are distinct. $u$ is regular, so $e_1' = e_1 \setminus \{u\}$ and $f_1'$ are disjoint. There can't be a regular path connecting $e_1', f_1'$ because that would be a hot path of $G_*^u$.

Applying lemma 8, in the contrapositive form, for the regular paths $\alpha' = (e_1', \ldots, e_n)$, $\beta' = (f_1', \ldots, f_m)$, we get that $e_n, f_m$ must overlap. Let $e_n = \{v, c, x\}$, $f_m = \{v, c, y\}$, and consider a third $u - v$ regular path $\gamma$ that is not going through $c$. There must be one, otherwise $c$ is a $u - v$ cutter. Assume wlog that $\gamma = (g_1, \ldots, g_k)$, $u \in g_1$, $v \in g_k$ is starting with an edge $g_1$other than $e_1$ (it can't be that $g_1 = e_1$ and $g_1 = f_1$ simultaneously) and apply the exact same argument for the paths $\alpha, \gamma$ to conclude $g_k = \{v, x, z\}$ for some $z$.

Consider the path $\alpha'' = (e_1', \ldots, e_{n-1})$. There are two cases: $x \in e_{n-1}$ or $c \in e_{n-1}$. Because of the symmetry between the triples $\beta, f_m, c$ and $\gamma, g_k, x$, we may assume wlog that $x \in e_{n-1}$, then apply lemma 8 for $\alpha'', \gamma'$ to get a hot path in $G_*^u$, contradicting $G_*^u \notin$ FAM1.                                                    □

## 3. Shortened Games are rather Short

The central theorem of this paper states that shortened games are rather short: If Maker moves first and has a winning strategy, then he can always win within two moves at most! (three moves, if we count breaker's move too). The proof by contradiction is long, and divided into numerous claims. Each claim settles a different scenario, or slowly adds up to known structure of the game we analyze. The reader is warned that the proof is not easily read, and familiarity with our terminology, notation system and previous lemmata is needed.

**Theorem 22.** *The maximal shortened depth of a winning game where Maker moves first is 3, and where Breaker moves first is 4.*

Assume for contradiction that the theorem is false.

*Claim* 23. There exists a game $G = (V, E, M)$ of shortened depth exactly 5.

*Proof.* Any game of shortened depth strictly higher than 5 has an winning sequence of optimal moves that be essence contains a game $G$ of shortened depth 5. By Corollary 19, Maker moves first in $G$.                                                    □

Let $H = (V, E)$ be the hypergraph of $G$, let $x_1, y_1, x_2, y_2, x_3$ be a sequence of optimal moves, starting with the first move $x_1$ by Maker. With this notation, the endgame we eventually reach is $G_{y_1,y_2}^{x_1,x_2,x_3} \in$ FAM2.

*Claim* 24. Neither of $G, G_{y_1}^{x_1}, G_{y_1,y_2}^{x_1,x_2}$ contains a hot path or a singleton edge.

If $\{u\} \in E_{y_1,y_2}^{x_1,x_2,x_3}$ then $\{u, x_3\} \in E_{y_1,y_2}^{x_1,x_2}$.

If $\alpha$ is a hot path of $G_{y_1,y_2}^{x_1,x_2,x_3}$ then at least one of the 2-edge ends of $\alpha$ is some $\{u, v\} \in E_{y_1,y_2}^{x_1,x_2,x_3}$ s.t. $\{u, v, x_3\}$ is a 3-edge of $G_{y_1,y_2}^{x_1,x_2}$.

*Proof.* $sdep(G), sdep\left(G_{y_1}^{x_1}\right), sdep\left(G_{y_1,y_2}^{x_1,x_2}\right) \geq 1$, hence $G, G_{y_1}^{x_1}, G_{y_1,y_2}^{x_1,x_2} \notin$ FAM1.

$\{u\}$ cannot be a singleton of $G_{y_1,y_2}^{x_1,x_2}$, so it must have only been formed in $G_{y_1,y_2}^{x_1,x_2,x_3}$ by withdrawing $x_3$ from $\{u, x_3\}$.

$\alpha$ cannot be present in $G_{y_1,y_2}^{x_1,x_2}$. All intermediate 3-edges of $\alpha$ are necessarily in $G_{y_1,y_2}^{x_1,x_2}$, so it must be (at least) one of the ends of $\alpha$ not in $G_{y_1,y_2}^{x_1,x_2}$, necessarily because of withdrawing $x_3$. □

*Claim* 25. There is a $x_2 - x_3$ regular path in $H_{y_1,y_2}^{x_1}$ (which is possibly a path of only one 3-edge $e$ s.t. $x_2, x_3 \in e$). $x_2$ is a regular vertex in $H_{y_1,y_2}^{x_1}$ and there is a $x_2 - x_3$ cutter in $H_{y_1,y_2}^{x_1}$.

*Proof.* If $x_2, x_3$ are not regularly connected, $x_2$ is not an optimal move of Maker, than can win by his second turn if he takes $x_3$ instead: all regular hot paths and singletons of $G_{y_1,y_2}^{x_1,x_2,x_3}$ are then present in $G_{y_1}^{x_1,x_3}$, so by lemma 20 the game terminated in $G_{y_1}^{x_1,x_3}$.

We get the second part of the claim by applying lemma 21 for $u = x_2, v = x_3$ and the game $G_{y_1,y_2}^{x_1,*}$, since $\left(G_{y_1,y_2}^{x_1,*}\right)_*^u = G_{y_1,y_2}^{x_1,x_2}$ is indeed not an endgame. □

*Claim* 26. All hot paths and singletons of $G_{y_1,y_2}^{x_1,x_2,x_3}$ are of the form $\gamma^{x_3}$ or $\alpha^{x_2,x_3}$, for a 3-uniform regular cycle $\gamma$ of $x_3$ in $E_{y_1,y_2,x_2}^{x_1}$ or a 3-uniform regular $x_2 - x_3$ path $\alpha$ in $G_{y_1,y_2}^{x_1,*}$. Moreover, there exist hot paths in $G_{y_1,y_2}^{x_1,x_2,x_3}$ of both forms.

*Proof.* By claim 24, any singleton or a hot path in $G_{y_1,y_2}^{x_1,x_2,x_3}$ not of those forms must be either a singleton $\{u\}$ s.t. $\{u, x_3\} \in E_{y_1,y_2}^{x_1,x_2}$ and $\{x_2, u, x_3\} \notin E_{y_1,y_2}^{x_1}$, or a hot path $\beta = (f_1, \ldots, f_m)$ s.t. $f_m = \{u, v\}$, $\{u, v, x_3\} \in E_{y_1,y_2}^{x_1,x_2}$, $f_1$ is a (virtual or proper) 2-edge and $f_1 \cup \{x_2\} \notin E_{y_1,y_2}^{x_1}$.

Let $\alpha = (e_1, \ldots e_n)$ be a regular $x_2 - x_3$ path of $H_{y_1,y_2}^{x_1}$ (existing by claim 25), $x_2 \in e_1, x_3 \in e_n$. In case of the singleton $\{u\}$, we may extend $\alpha$ with the edge $\{x_3, u\}$ to get a hot path $(e_1 \setminus \{x_2\}, \ldots, e_n, \{x_3, u\})$ of $G_{y_1,y_2}^{x_1,x_2}$, contradicting claim 24.

In case of the $\beta$ hot path, the conditions of lemma 8 with respected to the paths $\beta' = (f_1, \ldots, f_{m-1}, \{u, v, x_3\})$ and $\alpha' = (e_1 \setminus \{x_2\}, e_2, \ldots, e_n)$ must not be met, or else $G_{y_1,y_2}^{x_1,x_2}$ contains a hot path. Hence, either $f_1, e_1 \setminus \{x_2\}$ are not disjoint, or $\{u, v, x_3\}, e_n$ overlap.

From $f_1 \cup \{x_2\} \notin E_{y_1,y_2}^{x_1}$ we know $f_1 \neq e_1 \setminus \{x_2\}$, so if they are not disjoint they must form a v-shape in $G_{y_1,y_2}^{x_1,x_2}$, contradicting claim 24.

Say $e_n = \{w, v, x_3\}$ is overlapping $\{u, v, x_3\}$. The shared vertex of $f_{m-1}, f_m$ must be $u$ or $v$, and similarly shared vertex of $e_{n-1}, e_n$ must be $w$ of $v$. If it's $v$ in any of the pairs, then taking out the last edge of $\alpha'$ (resp. $\beta'$) if $e_{n-1} \cap e_n = \{v\}$ (resp. $f_{m-1} \cap f_m = \{v\}$) we may then use lemma 8 to get a hot path (and a contradiction).

We have proven that any pair of hot paths in $G_{y_1,y_2}^{x_1,x_2,x_3}$ s.t. one path is connecting an edge "touching" $x_2$ to an edge "touching" $x_3$ (e.g. the path $\alpha'$) and the other path is connecting an edge "not-touching" $x_2$ to an edge "touching" $x_3$ (e.g. the path $\beta'$), must share their last vertex (e.g. $v$). Hot paths of both types exist (the first type by claim 25, the second by our assumption), and hence the same vertex $v$ is essentially the end point "touching" $x_3$ in all hot paths of $G_{y_1,y_2}^{x_1,x_2,x_3}$, except for those of the form $\gamma^{x_3}$ for a 3-uniform linear cycle $\gamma$ of $G_{y_1,y_2}^{x_1,x_2}$ (e.g. hot paths "touching" $x_2$ in both ends).

$G_{y_1,y_2}^{x_1,x_2,x_3} \in$ FAM2, therefore $G_{y_1,y_2,v}^{x_1,x_2,x_3} \in$ FAM1. There are no singletons in $G_{y_1,y_2,v}^{x_1,x_2,x_3}$ ($\{v\}$ is the only possible singleton in $G_{y_1,y_2}^{x_1,x_2,x_3}$) so there must be a hot path of $G_{y_1,y_2,v}^{x_1,x_2,x_3}$, which is simply a hot path of $G_{y_1,y_2}^{x_1,x_2,x_3}$ not containing $v$, that must be then of the form $\gamma^{x_3}$, where $\gamma = (g_1, \ldots, g_k)$ is a linear 3-uniform cycle $\gamma$ in $G_{y_1,y_2}^{x_1,x_2}$ belonging to the vertex $x_3$.

$\gamma$ must be regular, or else it contains a virtual edge, and going along the cycle from that edge until we get to $x_3$ results in a hot path of a forbidden form. Any intersection of $\gamma$ with $\alpha'$ or $\beta'$ give rise (by lemma 8, connecting the two portions of the path and cycle) to a forbidden hot path.

Finally, we get a contradiction by showing $x_2$ to be a non-optimal move: If Maker takes $x_3$ instead, the resulting game $G_{y_1}^{x_1,x_3}$ is of FAM2, because it contains two hot paths $\beta'$ and $\gamma^{x_3}$ that are disjoint.

It remains to prove there are hots paths in $G_{y_1,y_2}^{x_1,x_2,x_3}$ in both of the forms $\alpha^{x_2,x_3}$ and $\gamma^{x_3}$. The existence of the $\alpha^{x_2,x_3}$ form is equivalent to $x_2, x_3$ being regularly connected in $H_{y_1,y_2}^{x_1}$ (claim 25). Since there exists a $x_2 - x_3$ cutter in $H_{y_1,y_2}^{x_1}$, not all hot paths of $G_{y_1,y_2}^{x_1,x_2,x_3}$ can be of the form $\alpha^{x_2,x_3}$ (else $G_{y_1,y_2}^{x_1,x_2,x_3} \notin$ FAM2), and any hot path not in the form of $\alpha^{x_2,x_3}$ is then in the $\gamma^{x_3}$ form. $\qquad\square$

*Claim* 27. If $x_1, x_2$ are not regularly connected in $H_{y_1}$ then we may assume wlog that $sdep\left(G_{y_1,x_1y_2}^{*,*,x_2}\right) \geq 3$.

*Reminder.* $G_{y_1,x_1y_2}^{*,*,x_2} = \left(V \setminus \{x_1, x_2, y_1, y_2\}, E_{y_1,x_1,y_2}^{x_2}, M\right)$ is the game we get if Maker is "passing" twice while Breaker do $y_1$ and $x_1$, following by Maker's $x_2$ then Breaker's $y_2$. Similarly, $G_{y_1,x_1}^{*,*}$ is the game $(V \setminus \{y_1, x_1\}, E_{y_1,x_1}, M)$, that we can think of as the game $G$ in which Breaker has been given two handicap moves $y_1, x_1$ before Maker is making the first turn (see §2.4 for the *-notation).

*Proof.* $sdep\left(G_{y_1,x_1}^{*,*}\right) \geq sdep\left(G\right) = 5$, so $sdep\left(G_{y_1,x_1}^{*,*,x_2}\right) \geq 4$ and there exists $y_2' \in V \setminus \{x_1, x_2, y_1\}$ s.t. $sdep\left(G_{y_1,x_1y_2'}^{*,*,x_2}\right) \geq 3$. We may assume wlog that Breaker is playing $y_2'$ as her second turn $y_2$ in $G$, because $y_2'$ serves as a possible optimal move of Breaker in $G_{y_1}^{x_1,x_2}$. All we need to do is prove $sdep\left(G_{y_1,y_2'}^{x_1,x_2}\right) \geq sdep\left(G_{y_1,y_2}^{x_1,x_2}\right)$, but $sdep\left(G_{y_1,y_2}^{x_1,x_2}\right) = 1$ so this is equivalent to $G_{y_1,y_2'}^{x_1,x_2} \notin$ FAM1.

Any singleton or regular hot path of $E_{y_1,y_2'}^{x_1,x_2}$ that is not present in $E_{y_1,x_1,y_2'}^{*,*,x_2}$ stems from a regular $x_1 - x_2$ path in $H_{y_1}$(assumed not to exist), $E_{y_1,x_1,y_2'}^{*,*,x_2}$ must include all singleton and regular hot paths of $E_{y_1,y_2'}^{x_1,x_2}$. By lemma 20, $G_{y_1,y_2'}^{x_1,x_2} \notin$ FAM1 follows from $E_{y_1,x_1,y_2'}^{*,*,x_2} \notin$ FAM1. $\qquad \square$

*Claim* 28. $x_1, x_2$ are regularly connected in $H_{y_1}$, and $y_2$ is a $x_1 - x_2$ cutter in $H_{y_1}$.

*Proof.* $H_{y_1,y_2}$ cannot contain a $x_1 - x_2$ regular path, or else $G_{y_1,y_2}^{x_1,x_2} \in$ FAM1. Thuse $y_2$ is a $x_1 - x_2$ cutter.

Assume $x_1, x_2$ are not regularly connected. By the previous claim, assume Breaker chooses $y_2$ s.t. $sdep\left(G_{y_1,x_1y_2}^{*,*,x_2}\right) \geq 3$, then $sdep\left(G_{y_1,x_1y_2}^{*,*,x_2,x_3}\right) \geq 2$ and there must be $y_3 \neq x_1, x_2, x_3, y_1, y_2$ s.t. $G_{y_1,x_1y_2,y_3}^{*,*,x_2,x_3} \notin$ FAM1. Consider the game $G_{y_1y_2,y_3}^{x_1,x_2,x_3}$, which must be in FAM1.

Any singleton $\{u\} \in E_{y_1y_2,y_3}^{x_1,x_2,x_3}$ cannot be in $G_{y_1,x_1y_2,y_3}^{*,*,x_2,x_3}$, so $\{x_1, u\} \in E_{y_1y_2,y_3}^{x_2,x_3}$. $\{u\}$ cannot be in $G_{y_1y_2}^{x_1,x_2}$ either (claim 24), so $\{u, x_3\} \in E_{y_1,y_2,y_3}^{x_2,x_3}$. The v-shape $\{x_1, u\}$, $\{u, x_3\}$ is impossible in $G$ (claim 24), so it must be that $\{x_1, u, x_3\} \in E$, contradicting claim 26.

Similar reasoning shows any hot path of $G_{y_1y_2,y_3}^{x_1,x_2,x_3}$ must have one end "touching" $x_1$ and the other end "touching" $x_3$, and is therefore contradicting 26. $\qquad \square$

*Claim* 29. We may assume wlog that $y_2$ is minimal with respect to $\prec_{x_2}$.

*Proof.* By the proof of claim 27, any vertex $y_2'$ s.t. $G_{y_1,y_2'}^{x_1,x_2} \notin$ FAM1 is a possible choice for Breaker's second move $y_2$ in $G$.[6] If $y_2$ is not minimal and there exists $y' \neq x_1, x_2, y_1, y_2$ s.t. $y_2' \prec_{x_2} y_2$, any $x_1 - x_2$ regular path contains $y_2$ and therefore contains $y_2'$ to, which means there are no hot paths or singleton in $G_{y_1,y_2'}^{x_1,x_2}$. $\qquad \square$

We are now ready to finish the theorem's proof. Consider the game $G_{y_1}^{x_1,x_3}$, that must not be in FAM2, or else $x_2$ is not an optimal move. There is a move $y_3 \neq x_1, x_3, y_1$ s.t. $G_{y_1,y_3}^{x_1,x_3} \notin$ FAM1. Hot paths in $G_{y_1}^{x_1,x_2,x_3}$ of the form $\gamma^{x_3}$ described in claim 26, cannot be present in $G_{y_1,y_3}^{x_1,x_3}$ and thus must include the vertex $y_3$.

$G_{y_1y_2,y_3}^{x_1,x_2,x_3} \in$ FAM1, and every hot path of $G_{y_1y_2,y_3}^{x_1,x_2,x_3}$ is also a hot path of $G_{y_1y_2}^{x_1,x_2,x_3}$, so by claim 26 the possible form are only $\gamma^{x_3}$ and $\alpha^{x_2,x_3}$ (described there). We have ruled out $\gamma^{x_3}$ for $G_{y_1,y_3}^{x_1,x_3}$, and since $\gamma^{x_3}$ is not "touching" $x_2$, it cannot be present in $G_{y_1y_2,y_3}^{x_1,x_2,x_3}$ either. Hence, there is a hot path $\alpha^{x_3}$ in $G_{y_1y_2,y_3}^{x_1,x_2,x_3}$ for a regular 3-uniform $x_3 - x_2$ path $\alpha = (e_1, \ldots, e)$, $x_3 \in e_1$, $x_2 \in e_n$ in $G_{y_1,*}^{x_1,*}$.

Let $\beta = (f_1, \ldots, f_m)$ be a $x_1 - x_2$ regular path in $H_{y_1}$ (with $x_1 \in f_1, x_2 \in f_m$) promised by claim 28. By claim 26there exists a hot path of the form $\gamma^{x_3}$, which must therefore contain $y_3$. If $\beta$ contains $y_3$, lemma 8 applies for $\beta, \gamma$ (since $\gamma$ is a regular cycle), resulting with a $x_1 - x_3$ regular path that yields a hot path touching $x_1$and $x_3$in the game $G_{y_1,y_3}^{x_1,x_3}$ which is not in FAM1. We therefore conclude that $y_3$is not present in either of $\alpha, \beta$.

---

[6]We have used the wlog-assumption purposed in lemma 27 only for the scenario (now ruled out) of $x_1, x_2$ not regularly connected, so there is no "wlog-abuse".

The requirements of lemma 8 hold for the paths $\alpha, \beta$ or any other pair of $x_1-x_2, x_3-x_2$ paths, then there exists a problematic $x_1-x_3$ regular path, contradicting $G_{y_1,y_3}^{x_1,x_3} \notin$ FAM1 again. The edges $e_1, f_1$ cannot overlap, or else replacing $e_1$ with $f_1$ in $\alpha$ yields a $x_1 - x_2$ path not including the $x_1 - x_2$ cutter $y_2$, contradicting claim 28.

Thus, the requirements of lemma 8 might fail for $\alpha, \beta$ either because $e_n, f_m$ overlap, or because the path $\alpha$ is not regular when the vertex $y_2$ is introduced back (by lemma 26, $\alpha$ is only promised to be regular in $G_{y_1,y_2}^{x_1,*}$). In both cases, an edge $y_2'$ in the overlapping pair of edges from $\alpha$ and $\beta$, is such that $y_2' \prec_{x_2} y_2$, contradicting lemma 29.[7]

## 4. The Polynomial Time Algorithm

**Theorem 30.** *There is a polynomial time algorithm that for every rank-3 Maker-Breaker game input $G$, calculates $sdep(G)$ and determines which player has a winning strategy in $G$.*

*Proof.* By theorem 22, Maker has a winning strategy in $G$ iff $sdep(G) < \infty$ iff $sdep(G) \leq 4$. Hence we are only interested in calculating $sdep(G)$, which the following straightforward algorithm does:

(1) Compute the game tree of (the shortened version of) $G$ to a maximal depth 4.
(2) Use a polynomial time algorithm (see fact 17) over all the leaves in the tree, to find those that are winning endgames, and set their $sdep$ field to be 0.
(3) Set the $sdep$ field of all other leaves to be $\infty$.
(4) Use the recursive rules of shortened-depth to calculate the $sdep$ field of all other games in the tree.
(5) Return the $sdep$ field of the tree's root $G$.

Note that the $sdep$ field calculated for a game $F \neq G$ other than the root might be greater than $sdep(F)$, but from theorem 22 it follows that the root's $sdep$ field is exactly $sdep(G)$.

4.1. **Running Time in the Random Access Model.** Let $G = (V, E, P)$, $n = |V|, m = |E|$. It is safe to assume $V = \bigcup_{e \in E} e$ since removing redundant vertices from $V$ keeps the game equivalent. Let $k$ be the number of virtual 2-edges in $G$, and remember that all virtual 2-edges of later games (i.e. games reachable from $G$ by a sequence of moves) are also virtual 2-edges of $G$. Each of the $m$ edges may contribute up to 3 virtual 2-edges and 3 new vertices, thus $n, k \leq 3m$ and $n, m = O(m)$.

The exact running time depends on the data structures used, but the algorithm essentially computes $O(n^4)$ games and determines for each whether it belongs to FAM2 (see lemma 18). Equivalently (see fact 17), the algorithm computes $O(n^5)$ games and determines for each whether it belongs to FAM1.

---

[7] While $y_2' \preceq_{x_2} y_2$ is immediate for the proper vertex $y_2'$, to rule out $y_2 \preceq_{x_2} y_2'$ is order to get $y_2' \prec_{x_2} y_2$, we need to keep in mind that there is not other pair of such $\alpha, \beta$ paths that meets the requirements of lemma 8.

Calculating a data structure for overlapping pairs of edges in $G$ and their virtual 2-edges intersections is possible in $O\left(m + n^2\right)$ time. Luckily, for each of the $O\left(n^5\right)$ games $G^{x_1,x_2}_{y_1,y_2,y_3}$ considered[8] it only takes $O\left(m\right)$ time to:

- Compute the game.
- Derive a specific data structure for overlapping and virtual edges in $G^{x_1,x_2}_{y_1,y_2,y_3}$ based on the data structure of $G$.
- Determine the existence of a singleton or a hot path (see lemma 14) in $G^{x_1,x_2}_{y_1,y_2,y_3}$.

Hence, the total time complexity is $O\left(n^5 \cdot m\right)$. In the (very probable, see §5.1) case $m = O\left(n^2\right)$ we get an upper bound $O\left(n^7\right)$ as a function of $n$ only.

## 5. Further Research

We would like to share some further research ideas that we find interesting.

### 5.1. **Extremal Combinatorics in Rank-3 Games.**
How many edges $m = |E|$ can we have, as a function of $n = |V|$, before any game $G = (V, E, P)$ must be a winning game?

There are many extremal combinatorics questions one could ask about Maker-Breaker games, and in particular rank-3 games. For the above question we get the lower bound $m \geq (n-2) \cdot \left\lfloor \frac{n}{2} \right\rfloor$ by taking

$$V = \{u_i | 1 \leq i \leq n\} \ , \ E = \left\{ \{u_{2k-1}, u_{2k}, u\} \, | u \neq u_{2k-1}, u_{2k} \text{ and } k \leq \left\lfloor \frac{n}{2} \right\rfloor \right\}$$

Is it maximal?

### 5.2. **Optimizing the Algorithm.**
Can it be sufficient to consider only a portion of the depth-4 game tree, and yet calculate $sdep\left(G\right)$ correctly? We believe so, because often there are moves dominated by (or equivalent to) other moves. Moreover, calculations that are repeated for multiple games in the tree game, might instead be performed only once in common. We believe the bounds $O\left(n^5 \cdot m\right)$ and $O\left(n^7\right)$ given in §4.1 are not tight.

### 5.3. **Solving Special Cases of Higher Rank Maker-Breaker Games in Polynomial Time.**
The complexity of 4-uniform Maker-Breaker games is PSPACE-complete[[me me me]], so assuming Assuming $P \neq PSPACE$ we cannot solve higher rank Maker-Breaker games efficiently. However, finding polynomial algorithms for higher rank games in special cases should be possible and interesting.

---

[8]Assuming Breaker moves first in $G$. If Maker moves first there are only $O\left(n^4\right)$ games, of the form $G^{x_1,x_2}_{y_1,y_2}$, the algorithm is dealing with.